\documentclass[11pt]{article}
\usepackage{moriond,epsfig,wrapfig}

\def\be{\begin{equation}}
\def\ee{\end{equation}}
\def\bea{\begin{eqnarray}}
\def\eea{\end{eqnarray}}

\begin{document}
\vspace*{4cm}
\title{STRUCTURE FUNCTIONS AND PARTON DISTRIBUTION FUNCTIONS AT THE HERA $ep$ COLLIDER}

\author{C.Targett-Adams, representing the ZEUS and H1 collaborations}

\address{Deutsches Elektronen-Synchrotron, Notkestrasse 85,\\22607 Hamburg, Germany}

\maketitle\abstracts{
Recent results on structure functions and parton distribution functions at
HERA are presented. Results on the measurements of the structure functions
$F_{2}$, $F_{L}$ and $F_{3}$ will be covered, as well as the results of the
most recent QCD fits to determine the PDFs of the proton.
}

\section{Introduction}
At HERA, 27.5$GeV$ electrons are collided with 920$GeV$ protons
\footnote{820$GeV$ before and including 1997}. ZEUS\cite{ZEUS} and H1\cite{H1} are the two general
purpose detectors which operate on the HERA ring. The $e^{\pm}p$ scattering
process is dominated by neutral current (NC) virtual $\gamma$ exchange with
charged current (CC) and NC $Z_{0}$ processes also providing a contribution at large
scales. Events for which the virtuality of the exchanged photon, $Q^{2} >
1GeV^{2}$ are classified as \emph{Deep Inelastic Scattering} (DIS). Events for
which $Q^{2}<1GeV^{2}$, whereby the exchanged photon is almost real, are
classified as \emph{Photoproduction} ($\gamma p$). Measurement of the
$e^{\pm}p$ differential cross section (with respect to the bjorken scaling
variable, $x$, and $Q^{2}$) yields information about the proton structure
functions $F_{2}$,$F_{L}$ and $F_{3}$. Further, performing QCD fits to HERA
cross section data allows the proton \emph{Parton Distribution Functions}
(PDFs) to be determined. Although the $Q^{2}$ dependence of the PDFs is
calculable using the DGLAP\cite{DGLAP1,DGLAP2,DGLAP3,DGLAP4} formalism of pQCD, the $x$ dependence has to be
determined empirically. HERA PDFs extrapolate into the LHC region and their
accurate determination are crucial to calculations of new physics measurements
at the LHC.

\section{Measurements of the structure functions: $F_{2}$,$F_{L}$ and $F_{3}$}

\subsection{Precision $F_{2}$ measurements} 
\begin{figure}
\begin{minipage}{3in}
\vspace{0.25in}
\hspace{-0.5in}
\epsfig{figure=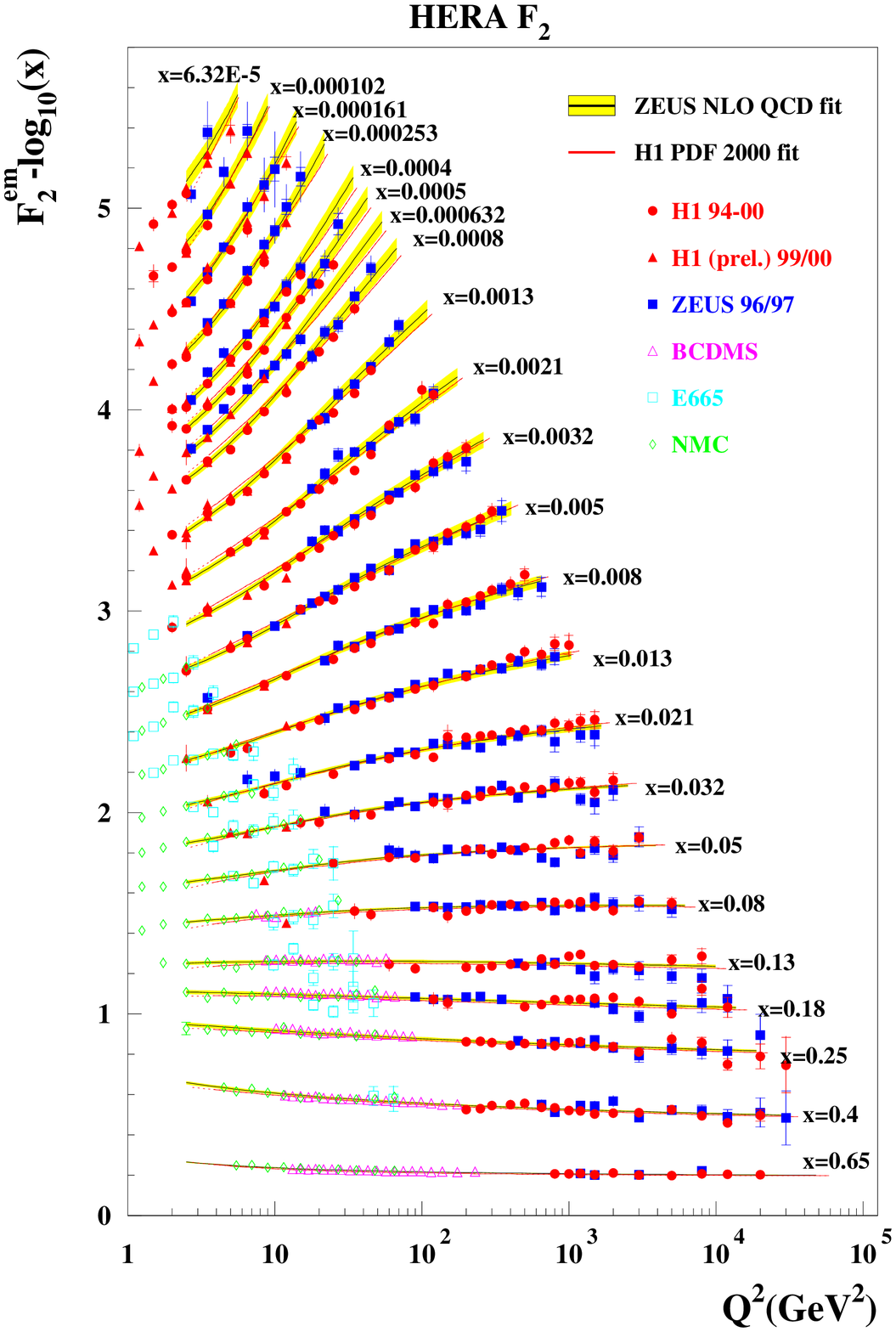,scale=0.45}
\caption{The most up-to-date measurements of $F_{2}$ from the H1 and ZEUS
  collaborations. Also shown are various heavy target $F_{2}$ results and the
  comparison of the measured structure functions with the H1 and ZEUS NLO PDF
  fits.}
\label{fig:F2}
\end{minipage}
\begin{minipage} {3in}
\begin{minipage}{3in}
\psfig{figure=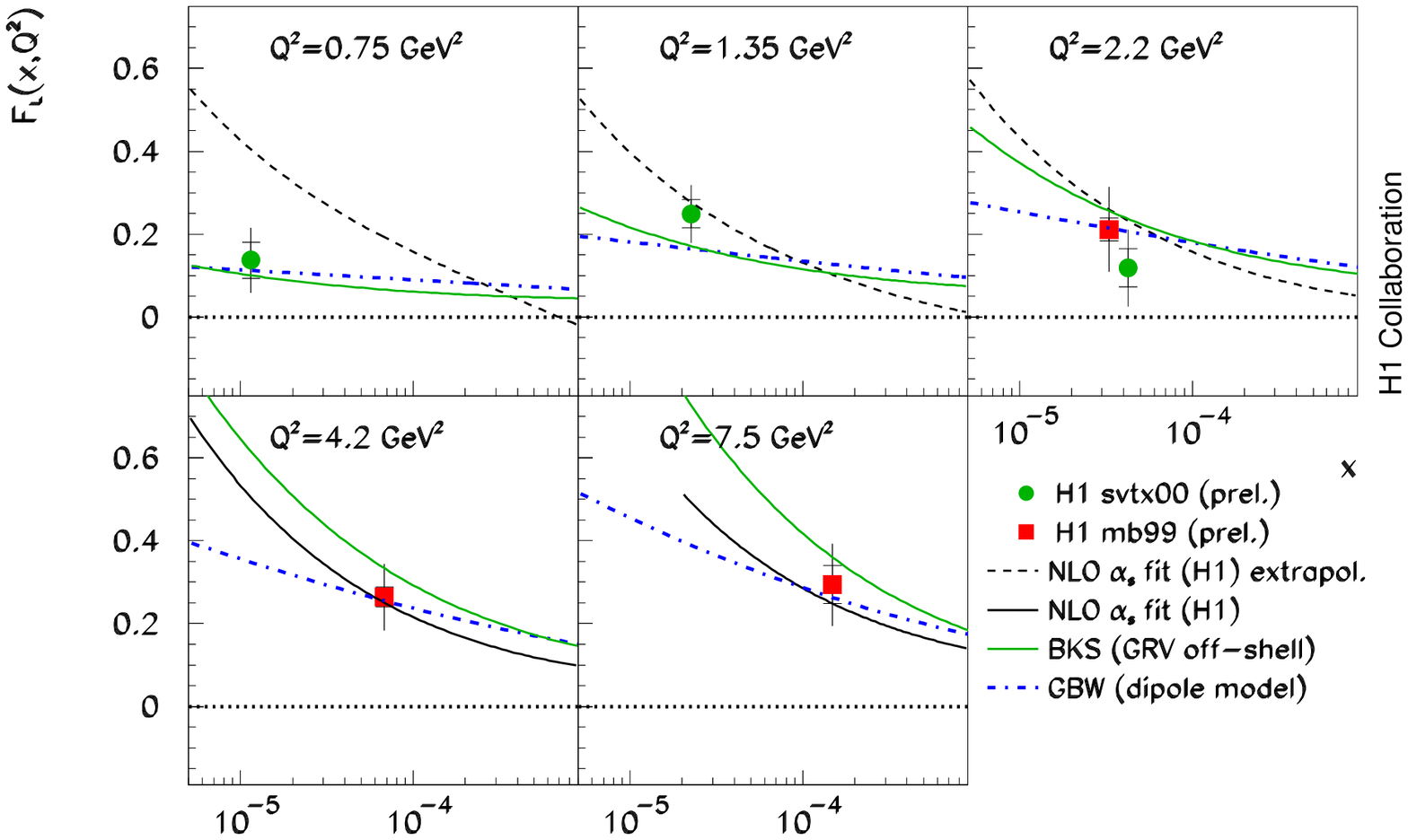,scale=0.40}
\caption{$F_{L}$ measurements from the H1 collaboration using their 'shape
  method' for extraction}
\label{fig:FL}
\end{minipage}
\begin{minipage}{3in}
\epsfig{figure=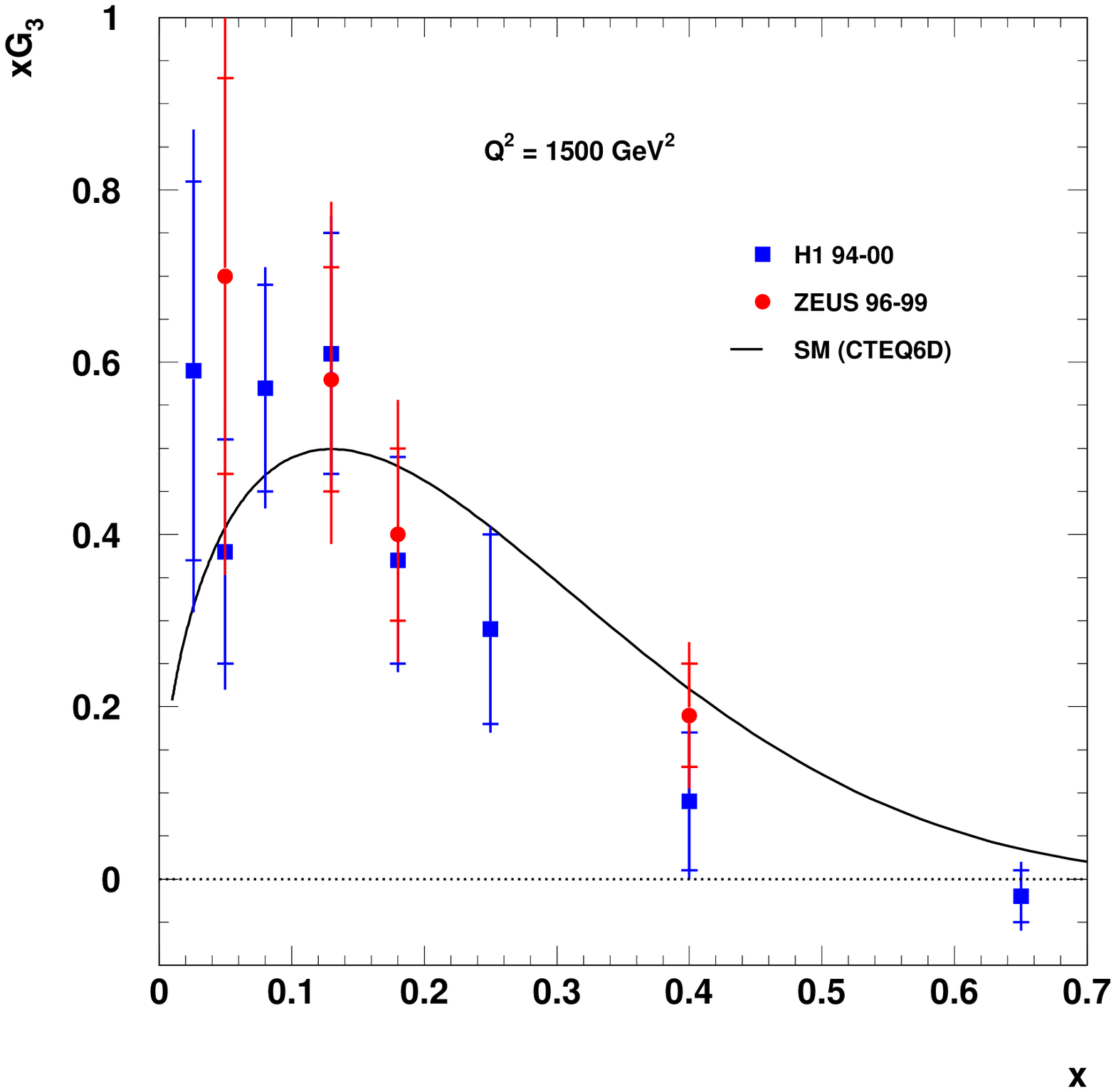,width=3in}
\vspace{-0.5in}
\caption{$F_{3}$ measurements from the H1 and ZEUS collaborations.}
\label{fig:F3}
\end{minipage}
\end{minipage}
\end{figure}

$F_{2}$ provides the dominant contribution to the neutral current $e^{\pm}p$
cross section. It is determined directly by measuring the so called reduced
cross section. Both ZEUS and H1 have measured $F_{2}$ to a typical accuracy of
$2-3\%$, across a broad kinematic range (approximately, $0.00001-1$ in x and
$1-30000GeV^{2}$ in $Q^{2}$) (see figure \ref{fig:F2})\cite{ref:F2}. $F_{2}$ is directly sensitive to the sum of all
the quark and anti-quark distributions inside the proton. It is however, only
indirectly sensitive to the gluon distribution via the phenomenon of scaling
violations. The $F_{2}$ data is well described by NLO pQCD.

\subsection{$F_{L}$ measurements}
Various measurements of $F_{L}$ have been made at HERA\cite{ref:FL1,ref:FL2}, predominantly by the
H1 collaboration. H1\cite{ref:FL1} have successfully used a parameterised fitting method,
called the 'shape method' to extract $F_{L}$ points, in a variety of $Q^{2}$
bins, from measurements of the reduced cross-section. Some of the results of
these measurements are shown in figure \ref{fig:FL}. At presently reached
accuracy the $F_{L}$ data is well described by NLO pQCD.

\subsection{$F_{3}$ measurements}
At high $Q^{2}$ the NC cross sections for electrons and positrons deviate. The
difference between the cross sections is proportional to $xF_{3}$ and allows a
direct measurement of this structure function to be made. The rather large
errors on the extracted $xF_{3}$ are dominated by the statistical error of the
electron sample. HERA has, to date, already built up significantly more
electron luminosity which will allow much more precise $xF_{3}$ measurements
to be made in the future. The $xF_{3}$ results from H1 and ZEUS are shown in
figure \ref{fig:F3}. 

\begin{figure}
\begin{minipage}{3in}
\vspace{-0.15in}
\hspace{-0.15in}
\epsfig{figure=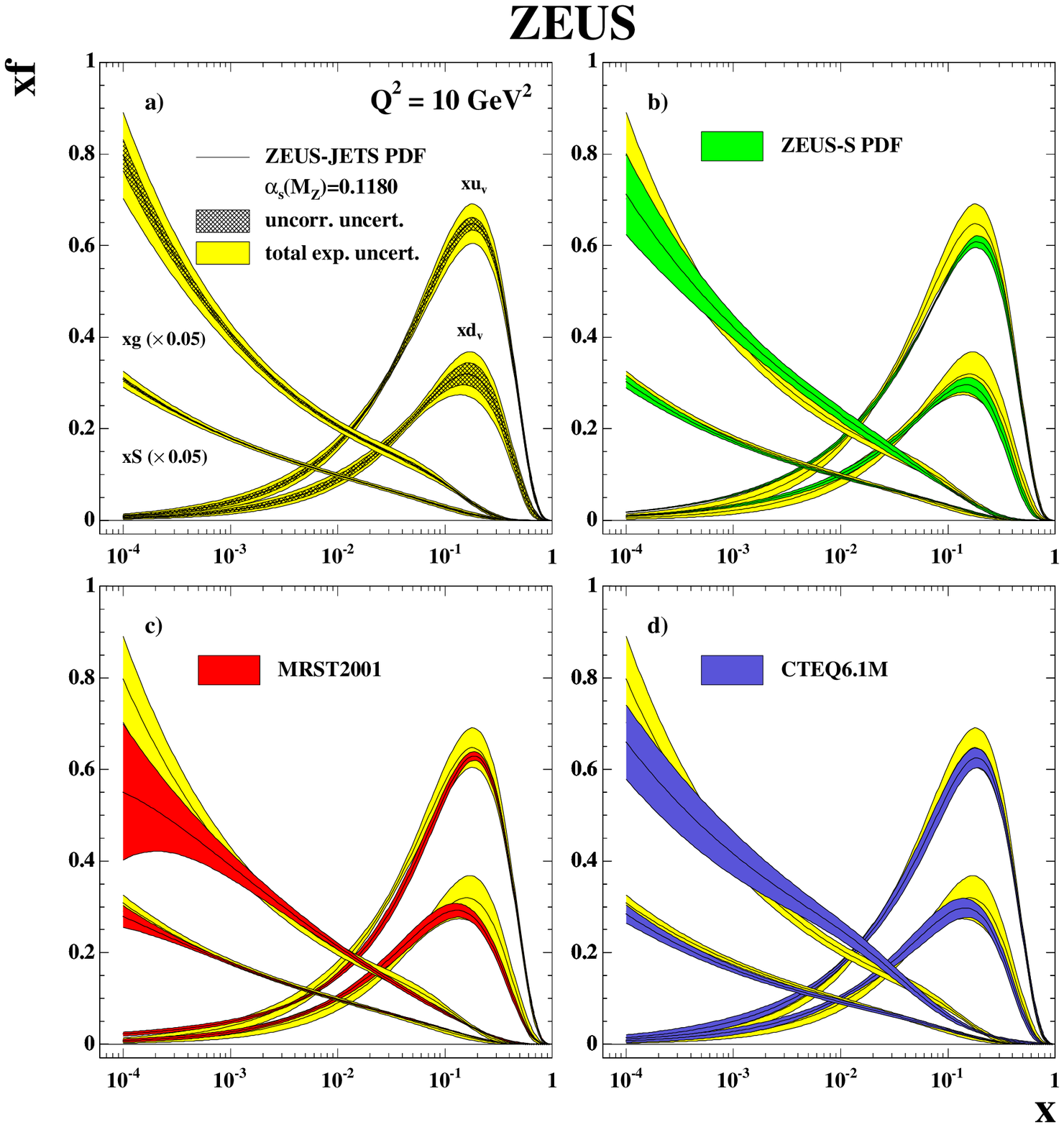,scale=0.4}
\caption{Comparison of the ZEUS-Jets PDFs with (a) The H1 fits (b) The
  previous ZEUS fits (c) The MRST fits (d) The CTEQ fits}
\label{fig:ZeusPdfs}
\end{minipage}
\begin{minipage}{3in}
\vspace{-0.25in}
\epsfig{figure=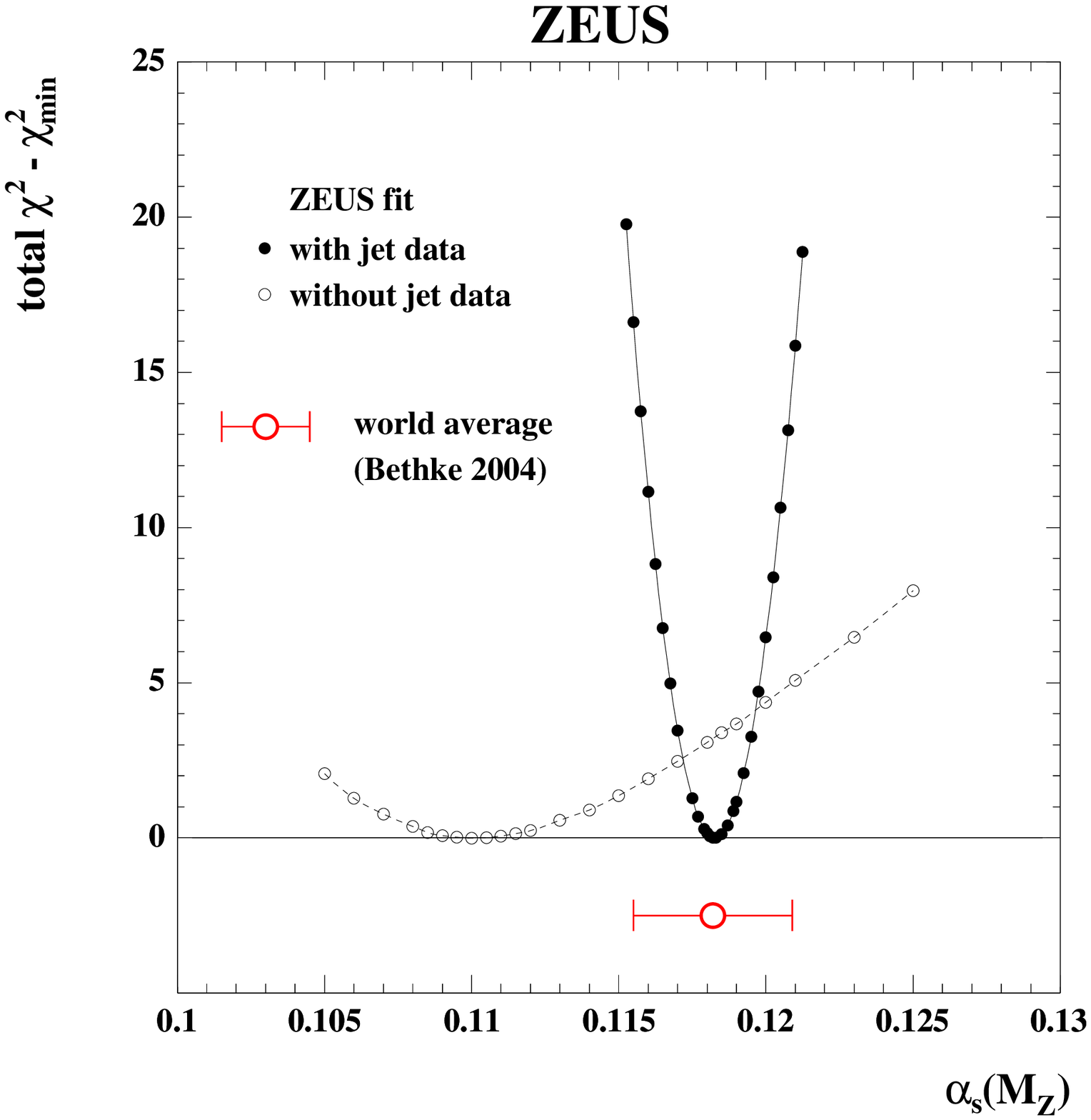,scale=0.42}
\vspace{-0.30in}
\hspace{0.3in}
\caption{$\chi^{2}$ profiles of the ZEUS fits with and without the inclusion
  of jet data in the fitted data set.}
\label{fig:chi2}
\end{minipage}
\end{figure}

\section{NLO QCD fits}
At ZEUS and H1, HERA data is used (along with external world data in the
latter case) to determine the proton PDFs. The general method involves
parametrising the PDFs at some input scale, $Q_{0}^{2}$. These are then
evolved to arbitrary $Q^{2}$ values using the DGLAP formalism and the
resulting PDFs are used to calculate cross sections and compared to data. The
starting parameters of the input distributions are then iteratively changed
until a best fit is found.

In the past, global world $F_{2}$ data has been used in the fits. Valence quark
distributions have been constrained by heavy target data ($\nu Fe$ and $\mu
D$). This has led to a poor $\alpha_{s}$ and gluon extraction as the inclusive
cross sections are only indirectly sensitive to the gluon via scaling
violations. 

\subsection{ZEUS-Jets fit}
The most recent fit to be performed at HERA is that which has been made by
ZEUS, the so called ZEUS-Jets fit\cite{ref:FITS1}. In the ZEUS-Jets fit no external
experimental data has been used, other than some minimal information which is
used to fix some of the initial parameters. High $Q^{2}$ NC and CC data
constrain the valence quark distributions, removing the necessity for fixed
\begin{wrapfigure}{l}{3in}
\hspace{-0.25in}
\includegraphics[scale=0.45]{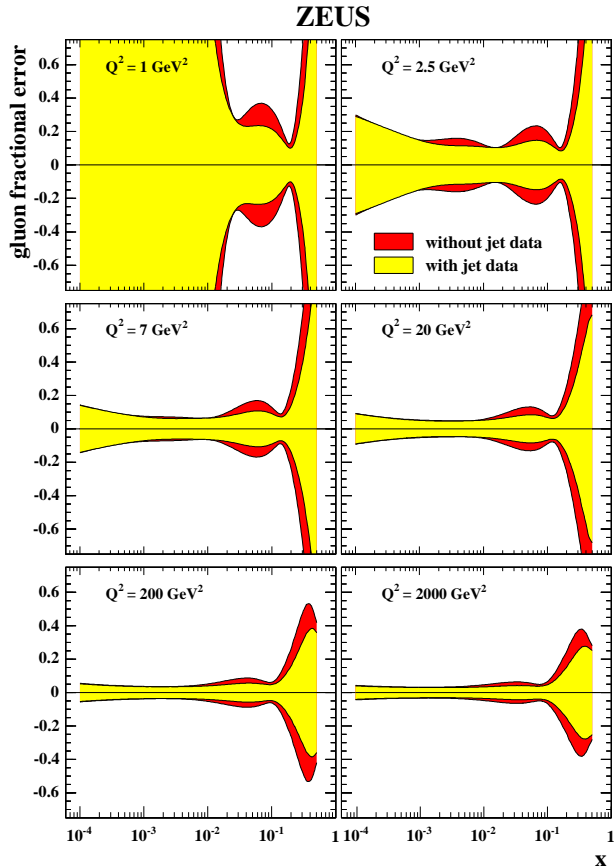}
\caption{Fractional uncertainty of the extracted gluon distribution from the
  ZEUS-Jets fit at a range of $Q^{2}$ values. The inclusion of jet data in the
  fitted data set has let to a clear improvement in the precision of the
  extracted PDF.}
\label{fig:gluon}
\end{wrapfigure}
target data to be included in the fits. Further, exclusive (jet) cross
sections have been included in the fitted data set for the first time. More
specifically, $\gamma p$ dijet data\cite{ref:DIJETS} and DIS inclusive jet data\cite{ref:INCJETS} from HERA
I\footnote{The running period up to 1997} have been included. This has led to a considerable improvement in the precision of the extracted gluon PDF. This is
illustrated in figure \ref{fig:gluon}. The PDFs, obtained from the ZEUS-Jets
fit, are shown in figure \ref{fig:ZeusPdfs}.


The values of $\alpha_{s}$ extracted from the ZEUS and H1 fits are given
below. 
\be
\alpha_{s} = 0.1150 \pm 0.0017 (exp) \pm 0.0007 (model)
\ee
for the most recent H1 fit and, 
\be
\alpha_{s} = 0.1183 \pm 0.0028 (exp) \pm 0.0008 (model)
\ee
for the ZEUS-Jets fit. Both values of $\alpha_{s}$ are characterised by a
scale uncertainty of about $\pm 0.005$.

Evidence of the effect that the jet data has in constraining the extracted
value of $\alpha_{s}$ in the ZEUS-Jets fit can be seen from figure
\ref{fig:chi2}. This figure shows the $\chi^{2}$ profile of the ZEUS fit when
the value of $\alpha_{s}$ is varied about its extracted value, with
and without the inclusion of jet data in the fitted data set. A much tighter
$\chi^{2}$ profile is seen with the inclusion of the jet data, corresponding
to a more constrained extracted $\alpha_{s}$.
\\
\section{Conclusion}

HERA continues to produce important research on proton structure and provide a
stringent testing ground for QCD. PDFs and $\alpha_{s}$ can now be extracted
with minimal data from external experiments. A rigorous inclusion of jet data
in to the fitted data sets has led to a significantly more precise gluon PDF
and $\alpha_{s}$ to be extracted from HERA data alone. More generally, the HERA
II program is well underway, which will give rise to evermore statistics
enriched data sets which in turn will give rise to both new and more precise
existing cross sections, thus paving the way for more accurate structure
function, PDF and $\alpha_{s}$ measurements.

\end{document}